\documentstyle[prb,aps,epsfig]{revtex}
\begin{document}
\baselineskip=12pt

\draft
\flushbottom
\twocolumn[\hsize\textwidth\columnwidth\hsize
\csname@twocolumnfalse\endcsname
\title {Doped magnetic moments in a disordered electron system:\\
insulator-metal transition, spin glass and `cmr' } 

\author{ Sanjeev Kumar and Pinaki Majumdar}

\address{ Harish-Chandra  Research Institute,\\
 Chhatnag Road, Jhusi, Allahabad 211 019, India }

\date{January 15, 2001}

\maketitle
\tightenlines
\widetext
\advance\leftskip by 57pt
\advance\rightskip by 57pt
\begin{abstract}

Recent experiments on the amorphous magnetic semiconductor Gd$_x$Si$_{1-x}$,
Phys. Rev. Lett.  {\bf 77}, 4652 (1996), {\it ibid} {\bf 83}, 2266 (1999), 
{\it ibid} {\bf 84}, 5411 (2000), {\it ibid} {\bf 85}, 848 (2000), have revealed  
an insulator-metal transition (i-m-t), as a function of doping and magnetic field, 
a spin glass state at low temperature, and colossal magnetoresistance close to the 
i-m-t.  There are also signatures of strong electron-electron interaction close to 
the i-m-t.  Motivated by these results we examine the role of doped magnetic moments 
in a strongly disordered electron system.  In this paper we study  a model of 
electrons coupled to structural disorder and (classical) magnetic moments, through
an essentially exact combination of  spin Monte Carlo and fermion exact 
diagonalisation.  Our  preliminary results, ignoring electron-electron interactions,
highlights the interplay of structural and magnetic `disorder' which  is primarily 
responsible for the observed features in magnetism  and transport.

\

\

\end{abstract}

]

\narrowtext
\tightenlines

\section{Introduction}

\subsection{Insulator-metal transition in doped 
semiconductors}

The role of disorder in electronic systems is an 
enduring 
theme in condensed matter physics
\cite{mottdavis,mottmit,andpap,leeram}.
Structural disorder leads to localisation of electronic states
and, in a three dimensional system,  all electronic states
would be localised \cite{andpap} if the disorder, `$\Delta$' say,
 were greater than a critical
value ($\Delta_c$). For weaker disorder,
$\Delta < \Delta_c$, only states {\it beyond} an energy $\epsilon_c(\Delta)$
of the band center, {\it i.e} in the band tails, are localised.
A  system is  metallic or insulating
depending on whether the Fermi level,
$\epsilon_F$, is above or below this `mobility edge' $\epsilon_c$.
Variation in  electron density,
and hence $\epsilon_F$,
 can drive a system through the i-m-t.
This is the basic scenario for the `Anderson transition' in
a non interacting electron system \cite{andpap,leeram}.

Doped semiconductors 
\cite{sarachikrev}
have served as a laboratory for
studying disorder effects
 since they allow systematic control
of the carrier density. 
The i-m-t has been studied in doped crystalline systems,
the most famous example being phosphorus doped silicon  
\cite{milligan}
(Si:P), 
and less extensively in amorphous semiconductors
\cite{bellloc}. In experimental systems, however,
the phenomena of localisation is complicated by
electron-electron (e-e) interaction effects which become
very pronounced close to the i-m-t \cite{altaron}.
There are clear signatures of interaction effects in the
density of states and the conductivity in amorphous 
systems close to the i-m-t.
The interplay of disorder and interaction effects near
the i-m-t  is not completely understood \cite{belkirk,mottand}
and continues to be an area of active research.

The introduction of magnetic moments in a `disordered' system
brings in new phenomena.
The  effects  
are  well understood in crystalline semiconductors \cite{kasurmp,wolff},
where the disorder is weak  and the relevant regime is of low
carrier density.
In these systems, electrons are trapped in the potential 
fluctuations and polarise
the magnetic moments in their neighbourhood. This leads to
a gain in exchange
energy, and tends to  enhance the localisation of the carrier.
The net localising effect in these systems, therefore,
 arises from a 
combination of $(i)$ 
on site disorder; a `single particle'
effect, and $(ii)$ `bound state' formation between the electron and the magnetic 
moments; a
`many body' effect.
The composite  object,  a (trapped) electron and
the polarised spins in its vicinity, is called a `spin polaron'.
Since the 
i-m-t in these systems occurs at 
a carrier density $\sim 10^{-5}$
(per unit cell),  the picture of non overlapping, 
uncorrelated,  spin polarons is adequate.
The interesting difference between these systems and
 their non magnetic
counterparts lies in the response to a magnetic field. An applied field
globally aligns the  magnetic moments,  reduces  the polaron binding
energy, and tunes  the  `magnetic component' of localisation. 
The change in 
activation energy (or mobility) of the carriers can  lead to 
{\it enormous magnetoresistance} \cite{vonMol83,ohno92}.
These systems have been known to exhibit  `colossal magnetoresistance'
(cmr) much before such
effects 
were  observed in the 
manganites. 
 
Recent experiments
\cite{hellman1,hellman2,hellman3,hellman4} 
 on amorphous $(a)$-GdSi reveal that doping 
magnetic moments in an {\it amorphous} system 
combines the richness,
and complication, of the  traditional i-m-t in amorphous semiconductors 
 \cite{bellloc}
 with the
physics of `cmr'. In addition 
a disordered magnetic state, a spin glass,
with rather unusual properties emerges at low temperature.
This combination of an i-m-t, in a high carrier density system,
with cmr, and a  spin glass ground state is probably
unique to GdSi.
We will discuss these experiments in the next section.
The essential phenomena in these
systems seems to be:
$(i)$ electron localisation due to the strong disorder in the
amorphous structure, $(ii)$ indirect exchange  interaction between
the doped  moments, mediated by the electrons, leading to
a spin glass state, $(iii)$
feedback of the spin background on the electronic system,
via `spin disorder', tending to enhance electron localisation,
$(iv)$ strong e-e interactions in the vicinity of
the i-m-t,  showing up as a correlation gap in the density of states, 
a $\sqrt T$ dependence of
the low temperature 
conductivity, and `local moment formation' in the electron system,
and, $(v)$
control of the  `spin 
disorder' by a magnetic field, $h$,  leading to large 
negative magnetoresistance when
the electron density 
is close to the critical density.

Unlike the case of crystalline magnetic semiconductors, 
where the picture of isolated  `bound magnetic polarons' \cite{kasurmp,wolff} 
seems to
suffice,
we do not have an understanding of a  {\it high density of electrons   
in the background of strong disorder, interacting with randomly located 
magnetic moments and with each other}. Scenarios in terms of an
Anderson transition, or isolated polarons, or model spin glasses, do
not suffice, and the interplay of these effects is what we study in this
paper.

While we are motivated by the phenomena in GdSi, which we take to be the
prototype amorphous magnetic semiconductor, 
our focus in this paper differs from the experiments in two respects.
$(i)$ 
In the experiments, discussed later, the tunable parameters
are the electron density $(n_{el})$
and the density of doped magnetic moments $(n_{sp})$.
Carrier density and `moment density', however, are not independent 
variables since both come from the doped Gd atoms. Thus, $n_{sp} \sim n_{el}$
and most of the data is for $n_{sp} = x \sim x_c$,
 with variations of a few percent. 
The `disorder' $(\Delta)$ due to the amorphous structure, or the coupling
$(J')$ between the electron and doped moments cannot be varied experimentally.
We will discuss a model which is expected to describe GdSi (see later) 
but not restrict ourself to parameters specific
to the real system right away.
The model has a large parameter space 
and  we will explore it gradually 
to locate some of the phenomena, i-m-t, spin glass and `cmr',
mentioned earlier.
We will present our results specific to  GdSi in a separate
publication \cite{gdsipap}. 
$(ii)$ The effects in GdSi arise from a combination of 
structural disorder,
electron-spin interaction, and e-e interaction.
All of them are important, in varying degrees, for the phenomena 
observed.
We will study a {\it model problem}
of electrons coupled to structural
disorder and magnetic moments, and ignore e-e interactions in
the present discussion. They are  important, 
as we will see in a detailed review of
the experiments,
but we have a non trivial problem even
without such interactions. As far as we know even this simplified
model has not been explored.  We will estimate and quantify  
the e-e interaction effects in our results specific to GdSi
\cite{gdsipap}.


Here is the outline of the text. In the next section we discuss the 
experimental results on amorphous ($a$)-GdSi 
in some detail,
to highlight the effects of doping magnetic moments into a
disordered system.
Following that
we define a model which
we believe contains most of the relevant physics.  We then  
discuss the approximations involved in `solving' for the properties
of this model and  present results on the magnetic properties, 
thermodynamics, and some simple limits for the
conductivity.
We conclude by indicating how electron-electron interaction 
effects can be included, approximately, within our scheme.
This would recover some of the features in the experimental data
which are not accessible in our `non-interacting' model.

\subsection{Magnetic moments in an amorphous background: experiments}

Treating GdSi as the `model' amorphous magnetic semiconductor
we review some of the effects which have been experimentally 
observed.
The measurements  have been made \cite{hellman1,hellman2,hellman3,hellman4}
 on $a$-GdSi 
and simultaneously on the  { \it non-magnetic analog}
 $a$-YSi, to clarify the
role of magnetic moments on the i-m-t. 
The results are broadly on $(i)$ the conductivity, $\sigma$, 
and the magnetoresistance,
$(ii)$ spectral properties/density of states 
(DOS), probed through tunneling,
 $(iii)$ 
thermodynamic properties: specific heat, $C_V$, and entropy, $S$, 
and $(iv)$
magnetic properties: the linear response susceptibility $\chi(T)$
and the magnetisation $m(h,T)$.

Let us begin by with the conductivity \cite{hellman1}. 
$(a).$ Both Y$_x$Si$_{1-x}$  and Gd$_x$Si$_{1-x}$
show an  insulator-metal transition 
as the doping, $x$, is increased 
across a critical value 
$x_c$. The critical doping 
required 
$(x_c \sim 14 \%)$ 
is slightly
greater in GdSi compared to YSi. 
$(b).$ For both GdSi and YSi,  in the doping
 range $0.17 > x > 0.11$, the conductivity 
has $d\sigma/{dT} >0$, {\it i.e} $d\rho/dT < 0$,
 at all $T$. This is true even of the systems 
which are `metallic' at low temperature.
$(c).$ For the metallic samples 
the low temperature conductivity
can be fitted to $\sigma(T) \sim \sigma_0 + \sigma_1 \sqrt{T} $.
$\sigma_0$ and $\sigma_1$ are constants and 
the $\sqrt{T}$ term arises  from Coulomb effects in a
disordered system \cite{altaron}. 
On  the insulating side $\sigma(T) \sim e^{-{\sqrt{T_0/T}}}$, indicating 
variable range hopping (VRH) in the presence of a soft Coulomb gap
\cite{efros}. 
$(d).$ All Gd based systems, in the doping range studied, show pronounced 
negative magnetoresistance
(MR). The MR increases on reducing $x$ (towards $x_c$), on 
decreasing $T$, and on increasing $h$.
YSi samples of comparable composition
show a significantly smaller, {\it positive}, MR.
$(e).$ 
The field dependence of conductivity 
is seen to arise principally from variations in $\sigma_0$ 
in the metallic phase 
\cite{hellmanssc},
and from the variation of $T_0$ in the insulator
\cite{hellmanprb}.
So, the dominant {\it temperature   dependence} of 
the conductivity in the metallic phase 
arises from Coulomb effects,
driving the $\sqrt T$, {\it but the MR arises almost entirely from
variations in the $T=0$ conductivity.}
$(f).$ For GdSi  samples with $x \lesssim x_c$ 
an applied magnetic field  can
drive the system metallic. 
Near this field driven i-m-t,  $\sigma_0$
varies as $(h-h_c)$, 
$h_c$ being the field  at
which the i-m-t occurs. The effects, $(a)$ and  $(d) - (f)$, 
hint at the crucial role 
of magnetic moments in determining  electronic transport.

The low energy DOS has been studied \cite{hellman4},
from measurement of the tunneling conductance, across
the field driven i-m-t. 
These measurements 
provide direct information on the 
evolution of the correlation gap \cite{altaron}
 (dip in DOS at $\epsilon_F$)
near the metal-insulator transition and the Coulomb gap 
\cite{efros} in 
the insulating phase.
The principal conclusion from these measurements
is that, for the field tuned transition,
the density of states 
at the Fermi level, $N(0)$, varies as $( h - h_c )^2$
while $\sigma_0(h) \sim (h - h_c)$. Therefore, 
the critical behaviour near the i-m-t follows
$N(0) \propto \sigma^2$, with both the DOS and the d.c conductivity
vanishing  at the transition.

The difference in the thermodynamic
properties of magnetic and non magnetic systems
again emphasises the role of the doped moments.
Even for disordered `non-magnetic' 
systems close to the metal-insulator transition the specific
heat has
a contribution arising from `local moments'  
\cite{paalnen}. This  arises because  electrons
can behave like {\it localised $S= 1/2$ objects}
 in the presence of
disorder and strong e-e interactions \cite{milovan}.
However, after `subtracting out' the specific heat of the 
non magnetic analog, one would expect the 
`high temperature' entropy of the magnetic system to 
correspond to 
$S^0(\infty) \sim n_{sp} ~ log (2 S + 1)$. 
Results on GdSi differ from this:
$(a).$ 
Specific heat measurements  indicate that at `high' temperature
$S_{mag}(T) = \int_0^T dT'C_V^{mag}/T'$ 
is  larger \cite{hellman2} than $S^0(\infty)$
by approximately  $50 \%$! 
The total magnetic
entropy tends to  
saturate, at this (larger) value
by $T \sim 60 - 70$ K.
A simple minded  estimate 
\cite{hellman2}
suggests 
that 
the additional entropy can be accounted for by 
$ \sim 2$
spin $1/2$ moments (conduction electrons) for each 
Gd moment.
$(b).$
$C_V^{mag}$ should vanish as $T \rightarrow 0$, however,
 down to $5$ K,  it shows
{\it no sign of a downturn} \cite{hellman2}. 
This suggests that the peak in  $C_V^{mag}$  occurs
somewhere below $ 5$ K. 
At high temperature the magnetic and non magnetic
systems have the same $C_V$.

Finally, the magnetism.
The doped Gd atoms  possess a moment $J= S= 7/2$, 
arising from the half-filled
$f$ shell.  The coupling between the randomly located 
moments  arises primarily through   
mediation by the conduction electrons.  
The overall magnetic effects also have contributions from the
`local moments' in the conduction electron system, alluded
to in the
previous paragraph.
The basic observations on the magnetism are:
$(a).$
Measurements at high fields, $h \sim 1$ T, show a magnetisation growing 
as $ \sim 1/T$, {\it i.e}, free moment like response. However,
measurements of $\chi$  at low field ($h \sim 10$ mT), 
reveals a transition \cite{hellman3} 
to a spin glass state at a temperature
$T_f \sim 1 - 6$ K depending on doping.
$(b).$
The freezing temperature increases from $T_f \sim 1$ K at
$x= 0.04$ to $T_f \sim 6$ K at $x \sim 0.20$.
The observation of the  characteristic 
\cite{sgrmp,sgfishhertz,sgmydosh} 
(logarithmic) frequency
dependent shift 
of $T_f$ confirms that these are indeed 
spin glasses.
$(c)$.
There is  
a distinct difference between
the  field cooled (FC) and zero field cooled (ZFC) susceptibilities 
$\chi(T)$; 
the FC susceptibility saturates to a higher value as $T \rightarrow 0$.
$(d).$
Fitting a Curie form to the susceptibility,
$\chi(x,T)= A(x)/(T - \theta(x))$, 
reveals that the effective moment, $\mu_{eff}$, which
can be extracted from $A$,  varies significantly with $x$.
Naively, this should have been just $ \sqrt{ S(S + 1)}$,
independent of $x$.
It is close to this expected (free moment)
 value for $x \approx x_c$ and falls
off on both the metallic and insulating sides.
$(e).$ The spin glass freezing does not have any signature 
in the temperature dependence of  transport
properties. The freezing itself is eliminated by fields $h \gtrsim 0.1$T.

These results suggest that
the magnetic state, which crucially affects electronic transport, is
itself determined by electron spin interaction, the background disorder,
and e-e effects.

Let us abstract our lessons from these effects, to start on a theory.
The {\it primary}
 effect in  amorphous magnetic systems 
is electron localisation due to disorder, 
either due to randomness in the amorphous structure or disorder in the 
spin configuration. 
To a first approximation, the i-m-t is driven by varying the mobility
edge $\epsilon_c$ across $\epsilon_F$. 
However, as the data indicates,  
the transition is not 
a simple `Anderson transition', driven by structural and spin disorder,
in a non-interacting electron system.
Electron-electron  interaction effects
are clearly in evidence in $(i)$ the temperature dependence of
$\sigma(T)$, $(ii)$ low energy features in the DOS, $N(\epsilon)$,
and, possibly also in 
$(iii)$ 
the `excess entropy', at high temperature, 
 and
$(iv)$ the non monotonic variation in 
$\mu_{eff}$ across the i-m-t. 

The transition, however,  is not driven 
by interaction effects {\it per se}.
Interactions become relevant near the transition, 
there the effects of disorder 
and interactions cannot be deconvolved,
but most of the phenomena occur {\it because the system is strongly
disordered. }
Since we will study a model without e-e effects,
note that many
of the experimental results {\it can be 
understood}, at least qualitatively,
even without such interaction. These include 
$(i)$ the {\it existence}
 of an i-m-t, from variation in the
`effective disorder' seen by the electron, $(ii)$ a magnetically
disordered ground state and the spin glass signatures, 
and, $(iii)$ the large MR from the field tuning of spin disorder.

This  problem, even without Coulomb effects, 
is non trivial because:
$(a)$ there is no simple
method for accessing transport properties in the
regime we are interested in (even if the `disorder'
were completely specified), and $(b)$ the `spin disorder' is {\it not
specified}, the magnetic state itself depends intimately on the electronic
spectrum and wavefunctions. 
This coupling is at the heart of the problem.

\section{The Model}

The model  for amorphous magnetic semiconductors
would have the form
\begin{eqnarray}
H = \sum_{ij} ( t_{ij, \sigma} c^{\dagger}_{i\sigma}c_{j \sigma} + h.c)
- \sum_i \mu n_i
&+& J'\sum_{\nu} {\bf \sigma}_{\nu}.{\bf S}_{\nu} \cr
&& + \sum_{ij} V_{ij}n_i n_j
\end{eqnarray}
where the label $i,j$ etc refer to the (non periodic) atomic 
locations ${\bf R}_i, {\bf R}_j$
etc. The labels $\nu$ refer to some set of positions $\{ {\bf r}_{\nu} \}$,
say, where the magnetic ions are located. The $t_{ij}$ would have a
distribution since the structure is non crystalline. The $V_{ij}$
stand for the Coulomb interaction.  
$\sigma_{\nu}$ is the electron spin operator
and 
we have assumed the local
electron-spin coupling, $J'$, to be site independent.

A simpler variant of this model, presumably with similar physics,
is the following:

\begin{eqnarray}
H = -t \sum_{\langle ij \rangle, \sigma } (  c^{\dagger}_{i \sigma}c_{j \sigma}  
+ &h&.c )
+ \sum_{i\sigma} (\epsilon_i - \mu) n_{i \sigma} \cr
+&J'&\sum_{\nu} {\bf \sigma}_{\nu}.{\bf S}_{\nu} 
+  \sum_{ij} V_{ij}n_i n_j
\end{eqnarray}
Here the $i,j$ refer to sites on a {\it periodic structure} and the $t$ now
refers to nearest neighbour hopping on that lattice. The `amorphous'
nature is incorporated via the random on site potential, $\epsilon_i$,
and the sites $\{ {\bf r}_{\nu} \}$ are some subset of the lattice points
$\{ {\bf R}_i \}$.
The parameters of the theory are $t$, $\langle \epsilon_i^2 \rangle
= \Delta^2$,
$J'$, $n_{el} = N_{el}/N $ and $n_{sp} = N_{spin}/N$. 
$N$ is the size of the system.
After normalising by $t$ there are two `coupling constants',
$\Delta/t$ and $J'/t$,  and two `densities':
four dimensionless system parameters in all. 
We will measure all energies in units of $t$ which,
from hereon, is set to $1$.
The Coulomb term, even if
included, does not involve a variable coupling constant. In addition  to
all these there are temperature and magnetic field as external variables.
In  what follows we will ignore the Coulomb term.
 We will also treat the core spin ${\bf S}_i$ as `classical', and 
study the magnetic properties, thermodynamics, and some
aspects of transport in this model. 

\section{Computational scheme }

Having defined the Hamiltonian we have to adopt a scheme for 
evaluating its properties. 
The problem involves strong disorder, in the variables $\{ \epsilon_i \}$, as
well as temperature (and field) dependent `disorder' in the spin
background $\{ {\bf S}_i \}$. Part of the problem is to determine the
distribution of spin configurations, 
appropriate to a given $T$ and $h$, the rest
to calculate and average electronic properties over the disorder
(on site  \&  spin). In the regime we are interested in, {\it
neither of these tasks is  analytically tractable.} Thankfully,
there is an essentially exact, and implementable,  numerical scheme 
which can be employed here. This is what we discuss next.

Consider the model for some 
 arbitrary coupling $J'$ and disorder $\Delta$.
The spins are considered to be classical (formally, $S \rightarrow \infty$).
The problem looks like  quadratic (non interacting) fermions
in the background of a 
spin configuration 
$\{ {\bf S}_i \}$. 
 The task is to  determine the 
appropriate background configuration(s), 
$\{ {\bf S}_i \}$, 
at a given temperature and specified values of the
electronic parameters. The problem can be set up formally as 
follows:
Consider a Hamiltonian $H = H_{el} + J'
\sum_i {\bf  {\sigma}}_i.{\bf S}_i$. Here
$H_{el}$ is a  quadratic fermion Hamiltonian
excluding terms involving the ${\bf S}_i$'s.
The partition function of this system is:  
\begin{equation}
Z = \int {\cal D}{\bf S}_i 
Tr_{c,c^{\dagger}}
e^{- \beta ( H_{el} + 
J'\sum_i
{\bf  {\sigma}}_i.{\bf S}_i )}
\end{equation}
We can formally trace over the fermions and write 
the partition function purely as an integral over spin 
configurations, {\it i.e},
\begin{equation}
Z = \int {\cal D}{\bf S}_i e^{ - \beta 
H_{eff}
\{ {\bf S}_i \} }
\end{equation}
where the effective `spin Hamiltonian', $H_{eff}$, satisfies
\begin{equation}
e^{ -\beta H_{eff}} = 
Tr_{c,c^{\dagger}}
e^{- \beta ( H_{el} + 
J'\sum_i
{\bf  {\sigma}}_i.{\bf S}_i )}
\end{equation}
It is obvious that the functional $H_{eff}
(\{ {\bf S}_i \} )$  is
just the fermion free energy $F_{el}$ in the background 
$\{ {\bf S}_i \} $.
 In general it depends on the full {\it  spin configuration}
$\{ {\bf S}_i \}$ and not just  pairwise interactions.
Our problem `reduces' to fermions in the 
background of some quenched disorder,  exchange coupled to 
spins {\it picked from a distribution}
$P( \{ {\bf S}_i \} ) \propto e^{ - \beta H_{eff} ( \{ {\bf S}_i \} ) }$.
To calculate physical properties involving the fermions, the
DOS or conductivity, say, we have to average over spin configurations
picked from the distribution $P$ appropriate to a given temperature.
The temperature dependence and the spin-spin correlations implicit
in $P$ indicate that the `disorder' arising from the $J'$ term
is very different from the quenched uncorrelated 
 disorder in the $\epsilon_i$'s.

Loosely,  the fermion properties can be calculated by
(weighted) average over probable spin configurations.
The weights themselves depend on a knowledge of the fermion
free energy in a given configuration! We have a coupled
problem here, and need to solve it self consistently.

\subsection{Exact enumeration}

We have written a formal prescription for the spin distribution: 
how do we proceed any further?
A computational scheme has been developed 
and explored over the last few  years for handling this kind
of problems
\cite{dagref1,caldref1}. Starting with some arbitrary configuration
$ \{ {\bf S}_i \}_0$, the 
 Metropolis algorithm is used for
updating the spin orientations and  generating new (acceptable)
configurations. The acceptance or rejection of a spin move, 
which depends on nearest neighbour orientations in short range
spin models, {\it now involves the diagonalisation of a $N \times N$
fermion Hamiltonian}. Unless some approximations can be
made, simulating such a model 
involves computational effort of
${\cal O}(N^4)$, where $N$ is the system size. 
A factor of  $N^3$ comes
from each diagonalisation (acceptance/rejection of a spin `move')
and another $N$ from the need to
`update' all $N$ spins in the system. This $N^4$ process has 
to be repeated $\sim 10^3 - 10^4$ times at each temperature
for equilibriation and averaging over equilibrium configurations.
At equilibrium, the spin configurations generated are a 
sampling of 
$P( \{ {\bf S}_i \} )$. Fermion averages calculated over these configurations
represent equilibrium average at that temperature.
While this scheme is `exact', and has no sign problems unlike fermion QMC,
the cost for handling large systems is still prohibitive. On a standard
DEC Alpha workstation one can handle systems $\sim 4 \times 4 \times 4$
{\it i.e} $N \sim 100$. Working out the temperature dependence of  $\chi$, 
say, for fixed electronic parameters,
involves about 12 hours of CPU time. 
The largest system studied \cite{dexsim}  is $6^3$ for the double exchange model.
To get a first impression of these complicated systems, the thermodynamics
and energetics 
of various phases can be reasonably studied for these small sizes. However,
where transport studies are of interest, and they are the most interesting
measurable property in these systems, the accessible sizes are much too
small for making useful statements (about $\sigma_{dc}$, say). 
This is because the mean free path $l_{mfp}$ in the low resistivity phases
is often 
larger than the system size $(l_{mfp} \gtrsim  L)$ and no useful
statements can be made about the conductvity of the thermodynamic system.
Some advances have been made by $(i)$  making the algorithms
more efficient that $N^4$, and $(ii)$ using `effective' spin models
for generating $P(\{ {\bf S}_i \})$,
rather than handling the full 
$H_{eff} (\{ {\bf S}_i \} )$. Algorithms in the first category are
rather complex and  still in
an experimental phase \cite{hybmc1,hybmc2}, 
while for approximations in the second category
there is often no obvious small parameter.
In this study we use an approximate scheme, described in the next section,
after  benchmarking it 
against small system simulations based on the exact scheme.

\subsection{Perturbative expansion  in $J'$}

While there is no obvious approximation for arbitrary values of the
quenched disorder $\Delta$ and the exchange coupling $J'$, 
the experimentally relevant regime in $J'$ (see later)
allows
a controlled approximation.
We have argued that 
$P( \{ {\bf S}_i \} ) \propto e^{ - \beta F_{el}  ( \{ {\bf S}_i \} ) }$.
Let us, for simplicity, look at $T=0$, in which case
$P( \{ {\bf S}_i \} ) \propto e^{ - \beta E_{el} ( \{ {\bf S}_i \} ) }$.
If the coupling $J'$ were small, in a sense to be quantified soon, 
we can try expanding the energy 
$ E_{el} ( \{ {\bf S}_i \} ) $, about the $J'=0$ limit, in
powers of $J'$. 
This is just quantum mechanical second order
perturbation theory, estimating the change in ground state
energy 
of the electrons due to coupling to an {\it arbitrary } spin configuration.
It is a variant of  the textbook `RKKY' 
argument\cite{kittelqts,fazekas}
and we briefly describe the steps below.

The reference system (at $J'=0$) is spin degenerate, so $\langle
\sigma_i^{\mu} \rangle_0 = 0$, for $\mu = x, y, z$,  and there is no 
${\cal O}(J')$ 
contribution
to the energy. The 
${\cal O}(J'^2 )$  term
is

\begin{equation}
\Delta E_0 ( \{ {\bf S}_i \} )
= \sum_{m \neq 0}
{ 
{\vert \langle
\Psi^N_0 \vert   
J'\sum_i {\bf  {\sigma}}_i.{\bf S}_i 
\vert
\Psi^N_m \rangle  \vert^2 } 
\over 
{E^N_0 - E^N_m} }  
\end{equation}
The $\vert \psi \rangle$'s are many particle wavefunctions 
(Slater determinants)
including the effect of on site disorder. The label $N$ refers to particle 
number here (not system size) and $m$ is the label for the state.
$E_0$ and $E_m$ refer to (unperturbed) ground state and excited state energies
of the $N$ electron system.
Writing out the energy change;
\begin{eqnarray}
\Delta &E_0& ( \{ {\bf S}_i \} ) \cr
\cr
& =& 
\sum_{m \neq 0}
{ {  
\langle
\Psi^N_0 \vert   
J'\sum_i {\bf  {\sigma}}_i.{\bf S}_i 
\vert
\Psi^N_m \rangle  
\langle 
\Psi^N_m \vert   
J'\sum_j {\bf  {\sigma}}_j.{\bf S}_j 
\vert
\Psi^N_0 \rangle  
 } 
\over 
{E^N_0 - E^N_m} }  \cr
& = &
J'^2 \sum_{i,j} \sum_{m \neq 0}
{ {  
\langle
\Psi^N_0 \vert   
 {\bf  {\sigma}}_i.{\bf S}_i 
\vert
\Psi^N_m \rangle  
\langle 
\Psi^N_m \vert   
 {\bf  {\sigma}}_j.{\bf S}_j 
\vert
\Psi^N_0 \rangle  
 } 
\over 
{E^N_0 - E^N_m} }  \cr
&=& \sum_{ij}J_{ij}{\bf S}_i.{\bf S}_j
\end{eqnarray}
where 
\begin{equation}
J_{ij}
=
J'^2 \sum_{m \neq 0}
{ {  
\langle
\Psi^N_0 \vert   
 {\bf  {\sigma}}_i^{\mu} 
\vert
\Psi^N_m \rangle  
\langle 
\Psi^N_m \vert   
 {\bf  {\sigma}}_j^{\mu} 
\vert
\Psi^N_0 \rangle  
 } 
\over 
{E^N_0 - E^N_m} }  
\end{equation}

The $\mu$ label in $\sigma$ could be $x$ or $y$ or $z$
(not summed over).
A large part of $\Delta E$ actually 
comes from the local, $S_i^2$, term but that
is a constant and does not affect the `cost' of rotating a spin with
respect to its neighbours.
In defining the effective spin Hamiltonian  we  
look only  at the $i \neq j$ terms in $\Delta E$.

To calculate $J_{ij}$,
suppose the single particle wavefunctions of the reference (disordered)
electron problem are of the form $\gamma^{\dagger}_{ \alpha \sigma}
\vert 0 \rangle = A^{\alpha}_ i c^{\dagger}_{i\sigma}\vert 0 \rangle$ 
and the inverse is $c^{\dagger}_{i \sigma}=B^i_{\alpha} \gamma^{\dagger}_{\alpha \sigma}$.
Since $A$ is a `rotation matrix' we will have $B= A^{-1}= A^{T*}$, 
The intermediate
state
$\vert \Psi^N_m \rangle $ is  created by a particle-hole creation operator,
of the form $\gamma^{\dagger}_{\alpha} \gamma_{\beta}$,  acting on  
$\vert \Psi^N_0 \rangle $ so it is indexed by two labels $\alpha, \beta, 
\alpha \neq \beta $.
Using this notation 
the exchange constants are specified as
\begin{equation}
J_{ij}= {J'^2 \over 4} \sum_{\alpha \neq \beta }
 \{ B^{\alpha}_iB^{\beta *}_i B^{\alpha *}_j B^{\beta }_j
+ c.c \}
( { n_{\alpha} - n_{\beta } \over {\xi_{\alpha} - \xi_{\beta}}})
\end{equation}

The $\xi$'s are single particle energies and the $n$ are Fermi factors.
It can be easily shown that $J_{ij}$ is 
$J'^2 \chi_{ij}(\omega =0)$ 
where $\chi_{ij}$ is  the spin susceptibility (response
function) of the electron system. For  electrons 
in free space  this is just
the Fourier transform of the standard `polarisability' $\chi_0({\bf q})$.
The (inverse) energy scale for $\chi_0$ is $ \sim N(\epsilon_F)$, {\it i.e}
we may write $\chi_0({\bf q}) = N(\epsilon_F) f_0({\bf q})$ where
$f_0$ is a dimensionless function ${\cal O}(1)$. Using this,
 the typical
magnitude of the (nearest neighbour) exchange constant is $\sim J'^2 N(
\epsilon_F)$.

Using similar arguments on a lattice, the exchange coupling
between nearest neighbour spins (on the lattice) will
  be $\sim J'^2/W$, since
$N(\epsilon_F) \sim 1/W$  where
$W$ is the bandwidth.
In the disordered electron system such a simple identification is no longer
possible but the typical nearest neighbour coupling will 
 still be
$\sim J'^2 N(\epsilon_F)$. 

It is apparent that if  $J'$  were small
there would be a {\it hierarchy} of energy scales
which may help us 
in analytically defining 
the effective 
spin Hamiltonian. If $J' \ll t$ then for the `clean' electron system 
we will have
$t \gg J' \gg J_{ij}$, while for the disordered problem we 
will have  
$\Delta \gtrsim t \gg J' \gg J_{ij}$. 
Experimentally the value of $J'$ is estimated \cite{hellman3}
 to be $ \lesssim 50$ K. Even with a modest $t \sim 1000$ K,
we would have $N(\epsilon_F)J' \sim 0.05 \ll 1$.
We will use a slightly larger $J'$, {\it i.e}  $J'/t \sim 0.2$,
and the 
smallness of $N(\epsilon_F)J'$ will  allow
us to work with a quadratic spin Hamiltonian.  Higher order
couplings, between three spins etc,
 will be smaller by this factor.

The primary scheme, as outlined at the beginning of this section,
 is to average electronic properties
by diagonalising the electronic part with spin configurations
chosen from the appropriate distribution. In our approximation,
the spin distribution is {\it specified} as 
$P \propto e^{- \beta H_{sp}}$, where $H_{sp} = \sum_{i \neq 
j} J_{ij} 
{\bf S}_i.{\bf S}_j$, 
and {\it 
does not
have to be numerically evaluated.}
However, we still have to do a Monte Carlo (MC), with long range spin-spin
interactions, to generate the equilibrium spin configurations.
 The problem involves the following
steps now, $(i)$ specify the exchange constants in terms 
of the spectrum and  eigenfunctions of the reference `site disorder'
problem, $(ii)$ perform classical Monte Carlo on the spin model
to generate a set of equilibrium configurations, in effect a 
sampling of $P( \{ {\bf S}_i \})$.
 $(iii)$ at a given temperature, after the spin system equilibriates,
diagonalise the electron problem with  some of the spin  configurations, and
average the computed correlation functions.
$(iv)$ since there is quenched disorder in the problem, repeat 
this entire
process for several realisations of $\{ \epsilon_i \}$ 
and average the observables.

If the  `rotation cost'  etc can be described by
a quadratic Hamiltonian then `polaronic' effects, which imply that spin rotation 
energy depends on the presence or absence of an electron  nearby, are {\it irrelevant}.
Polaronic effects  
may be relevant, 
{\it at large $J'$}, 
even in high carrier density 
systems \cite{dexsim}.  
The smallness of  $N(\epsilon_F)J'$ 
ensures that such effects are not relevant here.

\section{Results and Discussion}

We now present our results, starting with a comparison of the
`exact' and `approximate' schemes for our choice of $J'$.
We then discuss results on the d.c conductivity, in simple 
situations where the spin distribution does not have to
be evaluated through simulation. These data, nevertheless,
highlight the interplay of structural and spin disorder,
and provide an estimate of the obtainable MR. We finally examine
simulation results for magnetic moments in a (progressively) 
disordered 
electronic background and the spin glass signatures therein.

\subsection{Comparison of the exact and perturbative schemes}

Second order perturbation theory in $J'$ should be sensible
as long as $N(\epsilon_F)J' \ll 1$. 
We have argued that 
$J'$ is small for the systems of interest. 
There are two consequences of this. $(i)$
It is an enormous technical 
simplification 
because it `deconvolves' the magnetic problem from the
electronic one. For a given electron density, and a realisation
of disorder, we only need to compute the exchanges, $J_{ij}$, 
 once and
then work  with the resulting $H_{sp}$. Electronic properties can
be calculated by diagonalising the fermion Hamiltonian with
configurations $ \{ {\bf S}_i \}$ 
generated by equilibriating $H_{sp}$.
The computational cost reduces from $N^4$, for the exact
scheme, to $N^2$. In this $N^2$,
 one factor of $N$ is to evaluate 
the rotation cost for a single spin 
(in our long range model)
another  $N$ to update all the spins.
$(ii)$ Physically it implies that `polaronic' effects,
mentioned earlier are not relevant in this parameter regime.
The question of a `dense polaron system' does not arise here.

The single spin rotation cost is the central quantity in the
simulation.
To emphasise that this energy, at small
$J'$,  can be very accurately calculated from $H_{sp}$ we
have extensively studied this `rotation cost' under various conditions.
Fig.1 illustrates some of the cases. 
The energy cost
of rotation is the difference between 
the fermion
energies
$E \{ {\bf S}^1_i \}  - E \{ {\bf S}^0_i \} $, in
the 
exact scheme,
or 
$H_{sp} \{ {\bf S}^1_i \}  - H_{sp}  \{ {\bf S}^0_i \} $,
in the approximate one.
$ \{ {\bf S}^0_i \}$ and
$ \{ {\bf S}^1_i \}$ 
are two spin configurations connected by a single spin rotation 
$\delta {\bf S}_j$, say.
Fig.1 illustrates three cases. In each of them a reference
configuration 
$ \{ {\bf S}^0_i \}$ 
is chosen and $E$ and $H_{sp}$ are computed. A rotation is made,
changing  
$ \{ {\bf S}^0_i \}$ 
to
$ \{ {\bf S}^1_i \}$ 
and $E$ and $H_{sp}$ are computed on 
this state.
We plot $E_1 - E_0$ and $H_{sp}^1 - H_{sp}^0$ as a function of $J'$
for three reference configurations. 
The reference configuration 
$\{ {\bf S}_i^0 \}$ is random 
for the circle and 
square, it is ferromagnetic for the triangle.

\begin{figure}
\epsfxsize=7.5cm\epsfysize=8.0cm\epsfbox{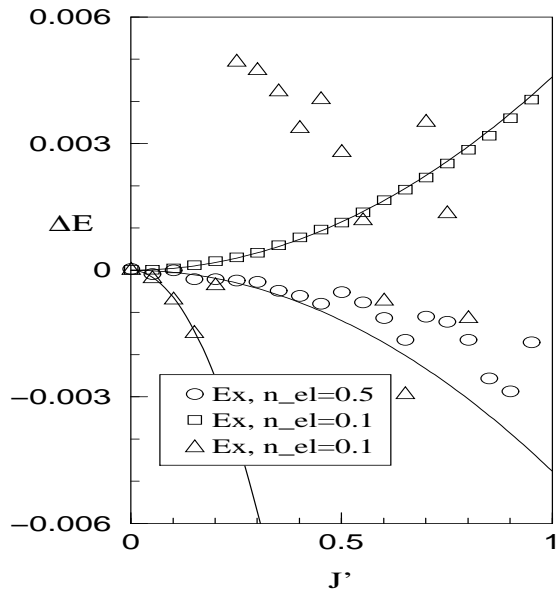}
\caption{Energy cost for rotation of a single spin: 
exact -vs- approximate schemes. 
The energy difference between two spin configurations is 
computed through direct diagonalisation (symbols) and by using $H_{sp}$
(lines), as a function of $J'$. The configurations are described in the text.
Disorder: $\Delta=8$, system size $14 \times 4 \times 4$. There is a spin at 
each site.}
\end{figure}

\begin{figure}
\epsfxsize=7cm\epsfysize=8.6cm\epsfbox{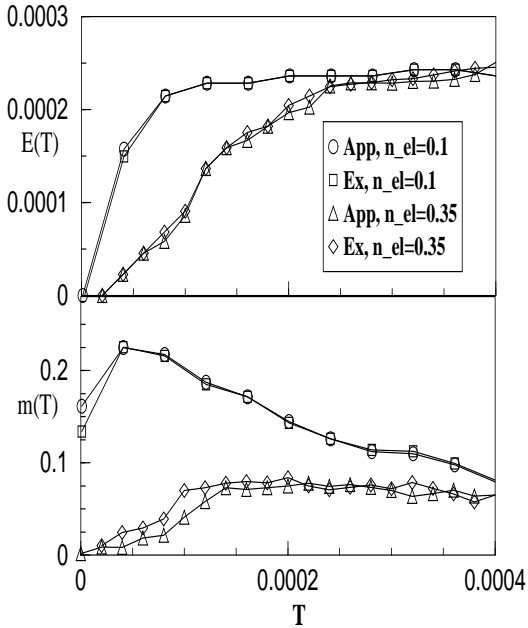}
\caption{
Exact simulation -vs- Monte Carlo with $H_{sp}$: clean system.
$J'=0.2$, $h= 10^{-5}$. 
The data is 
averaged over $1000$ configurations at each $T$, after
equilibriation over $1000$ steps. 
Total energy (top panel) and magnetisation (lower panel).
System size $4^3$. There is a spin at each site: $N_{sp}=N$.}
\end{figure}

The symbols are the `exact' cost,
the quadratic curves (firm lines) are
the 
approximate cost.
It is obvious 
that even in the worst case the `exact' and `approximate' results
match very well upto $J' \sim 0.2$. 

Since this configuration based testing cannot be exhaustive, we have
run the full simulation, 
using the exact scheme, and also done
MC with $H_{sp}$,  to check if the thermodynamics and
magnetism come out similar in the two approaches. 
We have used
$J'=0.2$ and studied a $4^3$  clean system (Fig.2) as well as a 
system with strong
disorder (Fig.3). We think the similarity is quite convincing.
We have also studied $J'=1$ (not shown) where the difference is quite
significant.

The principal conclusion from these tests is that we can reliably 
use an effective quadratic Hamiltonian
for the spins, and 
there are no 
polaronic effects in the parameter regime we are in.

\subsection{Results on transport}

To calculate the conductivity and
compare  with experimental data we would have to do a MC 
on the spin problem, having computed the $J_{ij}$, and use the
Kubo formula. 
The conductivity will have to be averaged over
equilibrium spin configurations. To compute the MR, {\it i.e}
$\sigma(h)$, the magnetic problem has to be redone in the 
presence of the field, and the cycle repeated. Even after

\begin{figure}
\epsfxsize=7.5cm\epsfysize=8.6cm\epsfbox{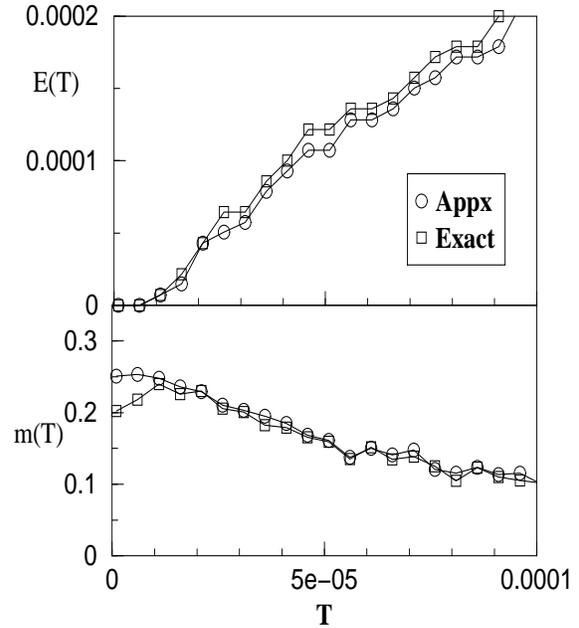}
\caption{
Exact simulation -vs- Monte Carlo with $H_{sp}$: disordered  system.
$\Delta=8$, $J'=0.2$, $n_{el}=0.1$, $h=10^{-6}$, size $4^3$, $N_{sp}=N$. 
The data is 
averaged over $1000$ configurations at each $T$, after
equilibriation over $1000$ steps. 
Energy (upper panel) and 
magnetisation (lower panel). 
}
\end{figure}

\noindent
the approximation we have made, this is a time consuming process.
In this paper we do not discuss those results, saving it for
the discussion specific to GdSi, but highlight some simple
limiting cases. 

We want to illustrate the interplay of
structural and spin disorder, and the difference in $\sigma$
between a spin polarised and random system.    

The conductivity is plotted in arbitrary units, {\it i.e} without 
putting in $e^2/\hbar$ and the lattice parameter.

\subsubsection{Effect of on site disorder}

First consider the case with only  `on site' disorder. This is
the traditional Anderson localisation problem.
We use the Kubo-Greenwood formula \cite{condcalc} to
calculate $\sigma$.  The data in Fig.4 displays the 
variation in conductivity (as $\epsilon_F$ is
varied)
and the DOS, with increasing disorder. 
$\sigma(\epsilon)$ is the variation in conductivity as the
electron density, or $\epsilon_F$, is varied. It is not
the `optical' conductivity.

The inset in Fig.4
shows the variation in the maximum conductivity in 
the band ($\sigma_{max}$ say)
which occurs at the band center here,
with increasing disorder.
Note the precipitious drop (even in this finite size calculation).
The most recent numerical calculations \cite{slevin}
put the Anderson transition in the 3d tight binding model at $
\Delta \sim 
16.5$.

To study localisation near the band edge 
 we need to follow the size dependence of
the conductivity for a fixed $\epsilon$ (or electron density).
If states at a certain energy, $\epsilon$, are localised then
$\sigma(\epsilon : L)$ will decrease with increasing $L$ 
(exponentially at large $L$). If the states are extended 

\begin{figure}
\epsfxsize=7cm\epsfysize=8.0cm\epsfbox{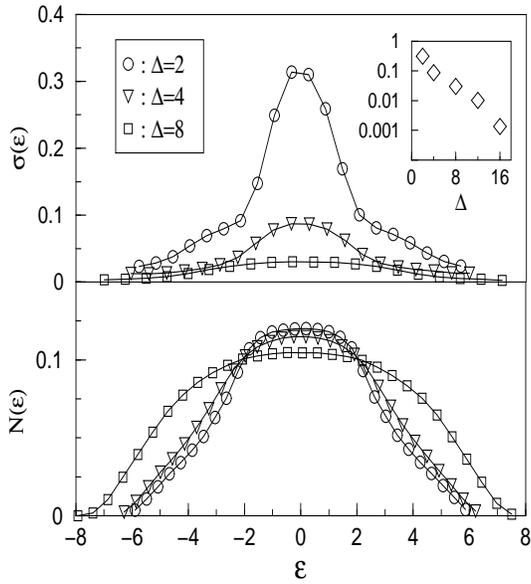}
\caption{Effect of on site disorder: conductivity (upper panel) and DOS (lower panel) 
in the presence of
 site disorder. System size $6 \times 6 \times 16$. The inset shows $\sigma_{max}$
the conductivity at the band center as a function of increasing disorder; notice that
the $\sigma$ scale is logarithmic. The data is averaged over $200-600$ realisations of
disorder depending on $\Delta$.}
\end{figure}

\noindent
then
$\sigma(\epsilon : L)$ will tend to a finite asymptote
as $L \rightarrow \infty$. Fig.5 shows $\sigma(\epsilon : L)$ 
for system size $6 \times 6 \times L$, with
 $L = 16 $ and $32$, 
for two strengths of disorder. The crossing point
is a crude measure of the mobility edge separating extended and 
localised states.
$\sigma(\epsilon : L)$ is 
approximately 

\begin{figure}
\epsfxsize=7cm\epsfysize=8.0cm\epsfbox{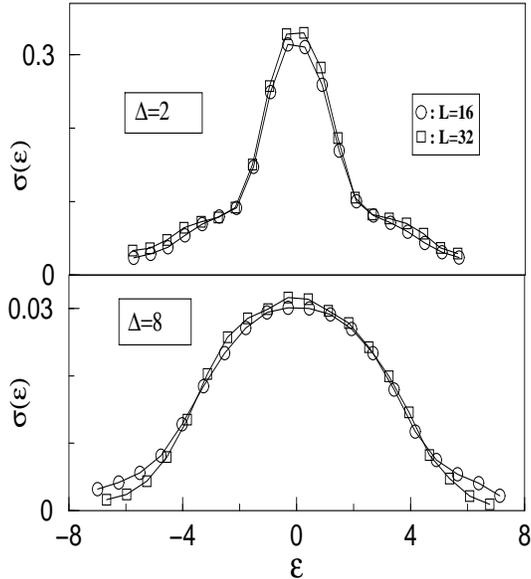}
\caption{System size dependence of the `conductivity', at two
values of disorder, $\Delta=2$ and $8$. The system sizes are $6 \times 6 \times 16$
and $6 \times 6 \times 32$. Average over $200 - 600$ realisations of disorder depending
on system size and $\Delta$. Notice the crossing of the curves in each panel.}
\end{figure}

\noindent
 independent of $L$ for states in the center of the band,
implying that our system size is large enough to provide 
an estimate of
the conductivity in the thermodynamic limit.

The purpose 
of presenting these results on the  very 
standard 
site disorder
problem is $(i)$ to
validate our conductivity calculation, and $(ii)$ to make an estimate
of the fraction of states localised as a function of disorder $\Delta$.
As a  crude estimate, $15 \%$ of the states (in the band tails) get
localised for $\Delta \sim 8$.
This will be useful for 
fixing parameter values in the GdSi problem.

\subsubsection{Effect of `spin disorder': uncorrelated spins}

Now consider how electron scattering off 
random magnetic moments affect the conductivity.
We consider a periodic array of spins with random orientations.
This mimics an (uncorrelated) paramagnetic state. There is no
on site disorder. The conductivity, computed by diagonalising
the electron problem in these random background,
is shown in Fig.6 
for three values 
of $J'$. The data is 
averaged over $100$ spin configurations for each $J'$.
Remember that to compute the conductivity  of
an electron system coupled to spins, the spin distribution has to
be {\it computed}, a non trivial task. Here we just {\it assumed}
an uncorrelated
random distribution.
This is expected to provide an estimate of $\sigma$ in the spin glass
phase (with $h=0$).
The 
inset in the top panel shows the reduction in the maximum conductivity
with increasing $J'$.

\begin{figure}
\epsfxsize=7cm\epsfysize=8cm\epsfbox{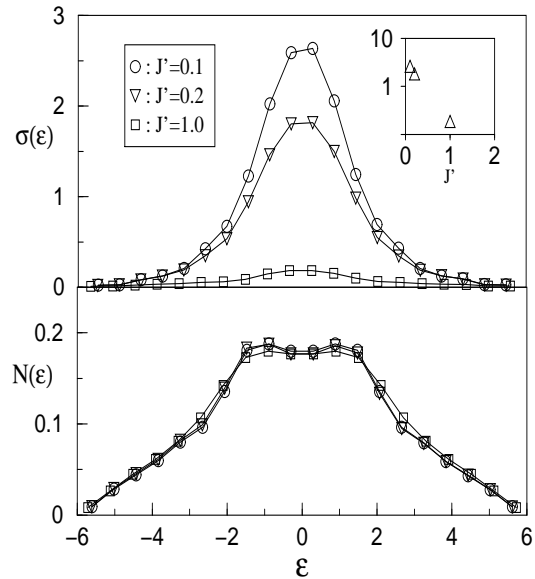}
\caption{Effect of `spin disorder': conductivity (upper panel) 
and DOS (lower panel) for electrons coupled
to randomly oriented spins. The exchange coupling is $J' = 0.1$, $0.2$ and $1.0$
and there is a spin at each site.
There is no on site disorder. Inset in the upper panel shows reduction in band
center conductivity with increasing $J'$. System size $6 \times 6 \times 16$.
Data averaged over $100$ spin configurations. }
\end{figure}

A similar problem, with $J'/t \rightarrow \infty$, has been studied 
earlier \cite{lanl} in the context of the double exchange model.
System size 
dependence of the conductivity in our results does not indicate
localisation in the band tails (upto errors $\sim 1 \%$).
In the $J' \rightarrow \infty$ problem \cite{lanl} it is known
that less than $1 \%$ of the states are localised
in the tail of the band.
Principally, in the small $J'$ limit in which we have
calculated the conductivity, electron 
coupling to `spin disorder'
leads to a  finite scattering rate  
$\sim {N(\epsilon_F)J'^2}$, and a small resistivity,
and not much by way of localisation.

We suspect that the mean free path at $J'=0.1$ and $0.2$ 
 is probably too large,
$l_{mfp} \gtrsim L$, to reproduce the `infinite volume'
conductivity.

\subsubsection{On site and `spin disorder': spin at each site}

The problem relevant for us has both on site and spin disorder.
Fig.7 shows the interplay of these two effects. The top curve
in the upper panel corresponds to on site disorder only, $\Delta=8$.
The lower curve (squares) shows the additional effect of random
spins, coupled to the electrons with $J'= 0.2$. If we focus on the
band center, for instance, the {\it resistivity}
increases from 
$\sim 1/{0.03}$ to $\sim 1/{0.015}$, {\it i.e},
from $\sim 33$ to $\sim 66$ (in arb units).
This {\it difference}, call it $\Delta \rho_{sp}$,
 arises due to  spin disorder. 
What would be the resistivity if  {\it only} this spin disorder
were operative?
This can be seen from the  lower 
panel in Fig.7. The `spin disorder only' resistivity, $\rho^0_{sp}$ is
$ \sim 0.5 \ll \Delta \rho_{sp}$, which is two orders of magnitude  
larger. 

\begin{figure}
\epsfxsize=7cm\epsfysize=8cm\epsfbox{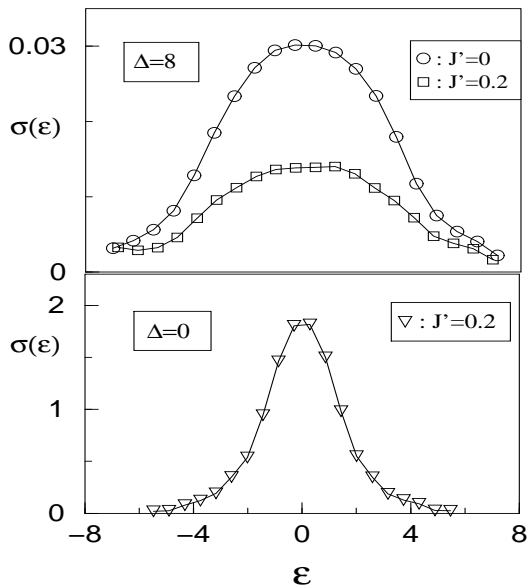}
\caption{
Interplay of on site disorder and spin disorder: conductivity
in the presence of a
random on site potential
and  randomly oriented spins {\it on a lattice} $(N_{sp}=N)$.
 Top panel shows data with on site disorder only
($\Delta = 8$ circles), and on site $+$ spin disorder ($\Delta =8, J'= 0.2$,
squares). The bottom panel shows $\sigma$ for $J'=0.2$, as in Fig. 6, without
any on site disorder. 
}
\end{figure}

We think that $\rho^0_{sp}$ is being underestimated, since 
$l_{mfp} \gtrsim L$. Nevertheless, even with a factor of $2$ error,
the difference between $\rho^0_{sp}$ and $\Delta \rho_{sp}$ is 
still significant.

In this disorder regime, the `on site' and `spin disorder' effects 
are far from additive:
 there is no Mathiessen's rule. This would of course be obvious 
close to the mobility edge, where spin disorder will lead to
localisation of additional 
states, but it is also prominently visible at the band center.

Within the accuracy of our calculation we
have not been able to see the shift in the mobility edge on 
introducing spin disorder (the shift, again,
 would be  $ \lesssim 1 \%$).
However, as noted,  even in the regime of {\it extended states}, 
there is a  dramatic increase in resistivity on introducing randomly
oriented spins, quite out of proportion to the small  
coupling, $J' =0.2$. 

The change in $\sigma$   (circles to squares, Fig.7) is, 
crudely, the difference between
YSi and GdSi: {\it i.e} the effect of on site disorder
-vs- on site $+$ spin disorder.
This also provides
an impression of the magnetoconductance obtainable, at
large field. The $J'=0$ curve on the top panel
is also the conductivity for {\it fully
polarised moments} (in which case the 
coupling becomes irrelevant). 

\subsubsection{On site and `spin disorder': dilute random array of spins}

Now for the final degree of realism.
Consider the effect of randomly {\it locating} the
spins (as in any amorphous structure) together with random orientation,
Fig.8.
There is also 
strong structural disorder in the background, $\Delta=8$.
There are three sources of randomness in 
this problem, in the site 
disorder $\{\epsilon_i \}$, the spin locations 

\begin{figure}
\epsfxsize=7cm\epsfysize=6cm\epsfbox{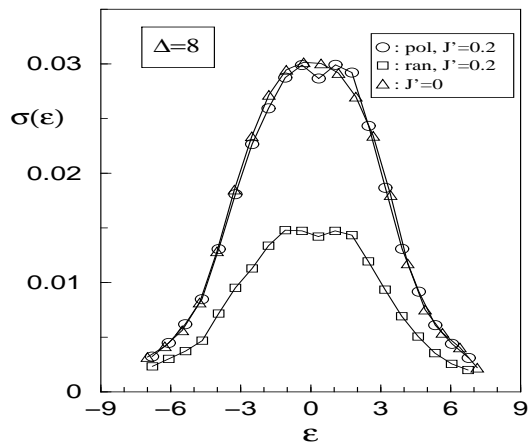}
\caption{
Interplay of on site disorder and spin disorder: conductivity
in the presence of a
random on site potential
and  {\it randomly located} array of spins.
The fraction of sites with spins is
$n_{sp}=0.3$, the  disorder is  $\Delta =8$ and  
the system size is $6 \times 6 \times 16$. 
The bottom curve (squares) 
corresponds to random orientation of the spins, one of the top curves 
(circles) corresponds to fully polarised spins, both with $J'=0.2$.
The third set (triangles) corresponds to on site disorder only
($\Delta=8$, $J'=0$) and is displayed for comparison. 
 }
\end{figure}

\noindent
$\{ {\bf R}_i \}$, 
and the orientations $\{ {\bf S}_i \}$. The
$\{ {\bf R}_i \}$'s were on a lattice in Fig.7. 
The data reveals that the effect of randomly locating  the moments
is of no consequence when the moments are polarised, $\sigma$ looks
as if there are no moments at all (top two curves). This is because
of the strong structural disorder already present. 
Polarised moments only `renormalise' the  
structural disorder from 
$\Delta$ to 
approximately 
$\sqrt { \Delta^2 + J'^2} \approx \Delta$.
Disorder in the {\it orientation}
 however has a strong effect, just
as for
periodically located spins, and all our arguments about MR, in
the last section, are applicable here as well. 
The difference between the top curve(s) and the lower curve would be
the `magnetoconductance' at strong field.
We are doing a more extensive calculation to quantify
the localisation 
effects near the mobility edge.

\subsection{Magnetic moments in the electron system: effect of increasing
disorder}

The previous section set
 out some of the limits for the conductivity,
arising from site disorder and randomly oriented or fully polarised
spins. For these results to have any relevance the actual magnetic
state in the  system should bear some resemblance 
to the random  magnetic state we used.
In this section we study the magnetic state that arises for
moments coupled to electrons in a disordered environment,
and the conditions under which a `spin glass' state can arise.
We first consider a periodic system, {\it i.e} one with a spin on
each site of the lattice
and study the thermodynamic and magnetic properties as the 
disorder, $\Delta$, is increased.

\subsubsection{Periodic array of spins}

The magnetic properties are studied via Monte Carlo on the
Hamiltonian $H_{sp} = \sum_{i \neq j} J_{ij} {\bf S}_i . {\bf S}_j
$ as discussed earlier. The $J_{ij}$ depend on electron density,
$J'$,  and {\it the specific realisation of disorder in } $\epsilon_i$'s.
For a clean system, $\Delta = 0$, the bonds $J_{ij}$ depend only on
the separation $i - j$ irrespective of the location of the two sites.
More concretely, when using periodic boundary conditions (PBC), 
all nearest neighbour bonds
will have the same value $J_1$, next nearest neighbour bonds will all be
$J_2$, and so on.

The bond {\it  distribution}
for nearest neighbour bonds, call it $P_1(J)$ will be a $\delta$ function.
Similarly $P_2(J)$, for next nearest neighbours, $P_3(J)$, for
third neighbours,   will all
be $\delta $ functions. 
These 
bond distributions broaden out 
 when the electron system is
disordered. 

The bond distribution, upto third neighbour, is shown on Fig.9 for
a $6^3$ system, with PBC, for $\Delta = 2$, $4$, and $8$. The
`spikes' are the `distribution' for $\Delta = 0$. Notice that 
although the first neighbour coupling remains ferromagnetic from
$\Delta = 0$ to $8$, the farther neighbour couplings are 
antiferromagnetic and not much weaker than the first neighbour 
 coupling 
(at strong disorder). 

\begin{figure}
\epsfxsize=7cm\epsfysize=7cm\epsfbox{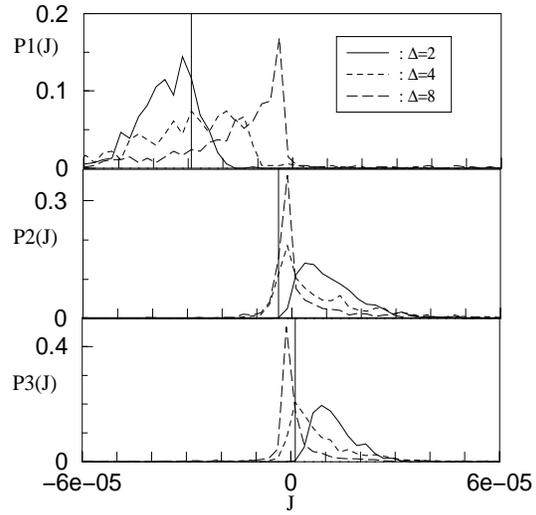}
\caption{Distribution of bonds: nearest neighbour, $P_1(J)$,
next neighbour, $P_2(J)$, and third neighbour,  $P_3(J)$, as a
function of increasing disorder. The spins are on a $6^3$ lattice,
the electron density is $0.1$ and $J'=0.2$. The spikes ($\delta$ functions)
correspond to the clean case $(\Delta=0)$. The bonds are calculated 
for  a
single realisation of disorder for each $\Delta$. 
The normalising factor for
all the distributions is $4 \times 10^5$.}
\end{figure}

The clean system is a model case of `RKKY', with some difference arising 
from the finite size and  `lattice' nature of the system. The strongly 
disordered
systems also seem to have reasonably long range interactions
\cite{rkkydisord},
but frustrating 
instead of regular as in the RKKY case. 

The next figure, Fig.10, shows the result of simulations on $H_{sp}$.
The internal energy $E(T)$ and the weak field 
magnetisation $m(T)$ are 
monitored. 
The data is averaged 

\begin{figure}
\epsfxsize=7.5cm\epsfysize=7.0cm\epsfbox{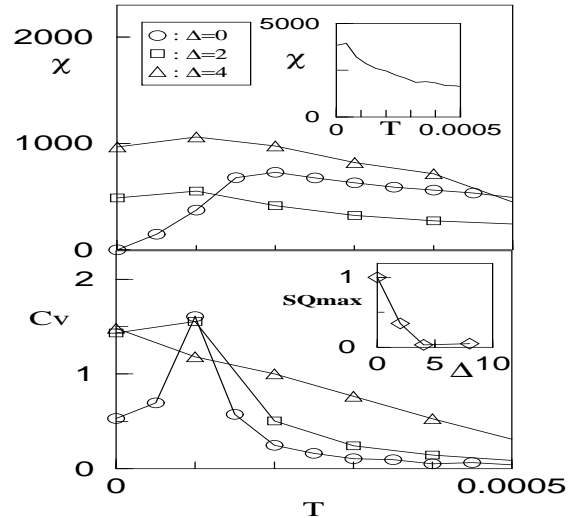}
\caption{
Susceptibility, specific heat, 
and structure factor in the
bond disordered periodic magnetic system.
 Upper panel, susceptibility $(\chi)$,
lower panel, specific heat $(C_V)$, for increasing disorder,
$\Delta =0$, $2$, $4$. The system size $6^3$, $J' = 0.2$,  and $n_{el} = 0.1$.
The upper inset shows $\chi$ for $\Delta=8$, and the lower panel 
shows the evolution of the {\it peak amplitude}
 in the structure factor,
$S({\bf q})$, with increasing disorder. }
\end{figure}

\noindent
over $\sim 1000$ equilibrium 
configurations at each temperature and then over $2 - 8$ realisations
of disorder, depending on $\Delta$. 
The following are the important features in the data: $(i)$
All the way from $\Delta =0$ to $\Delta= 8$ there is a cusp in
the susceptibility. The temperature at which it occurs reduces
from $\sim 1.8 \times 
10^{-4}$ at $\Delta = 0$, to $0.5 \times 10^{-4}$ at
$\Delta = 8$ (see inset in top panel, Fig.10). 
$(ii)$
The magnetic specific heat 
has a well defined peak at $\Delta = 0$, 
which  gets
broadened with increasing disorder. The
specific heat is larger 
in the bond disordered systems at lower temperature. It is
useful to remember that these are {\it classical} spins, so the
$C_V$ remains finite at $T=0$, $(iii)$ the feature in $\chi$ and
$C_V$ suggests an ordering transition in the clean system, while 
the nature of the low temperature state in the disordered systems is
not obvious purely from $C_V$ and $\chi$.
 Since the spins are on a periodic structure, we calculated
the structure factor $S({\bf q})$ from the ground state configuration
in each case (inset, lower panel). 
$(iii)$ The clean system has a single peak in
$S({\bf q})$  at ${\bf q} =
0, 0, \pi$, of magnitude $ \sim {\cal O}(N^2)$. 
The system is ordered as expected.
 A reduced   peak survives 
at the same 
${\bf q}$ for $\Delta =2$, see Fig.10, but disappears at 
\noindent
larger disorder.
The feature in $\chi$, along with  the absence of any long range order,
suggests that the strongly disordered systems are actually spin glasses.
We are in the process of computing the Edwards-Anderson order parameter
and checking the system size dependence of our results
on $\chi$ and $S({\bf q})$.

\subsubsection{Dilute random array of spins}

Finally, let us look at the magnetism in a site diluted clean
system. 
This would be the canonical RKKY model 

\begin{figure}
\epsfxsize=7cm\epsfysize=7cm\epsfbox{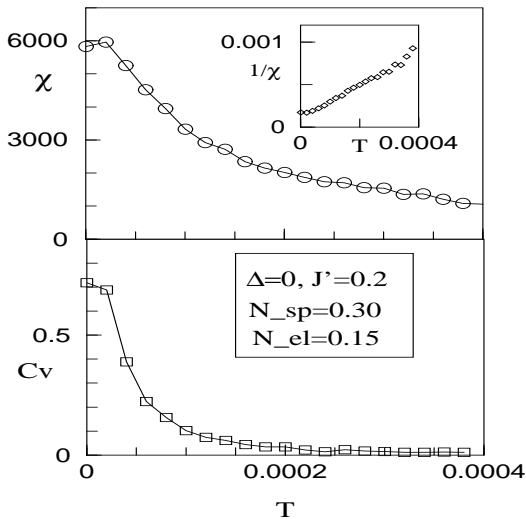}
\caption{
Susceptibility and  specific heat
in the {\it clean} diluted spin system. The
spins are distributed on 
randomly chosen 
$30 \%$  
of the  sites
in a $8^3$ system. The electron density is $0.15$, $J'=0.2$ and 
$\Delta=0$. Applied field $h = 10^{-5}$. The
data is averaged over $1000$ MC steps at each temperature
 after equilibriation over 1000
steps.  }
\end{figure}

\noindent
except that the real
exchange in a lattice model looks somewhat different from the
simple $r^{-3} cos(2k_Fr)$ form usually assumed in simulating
such models.
We show the $\chi$ and $C_V$ in such a system, an $8^3$ lattice
with spins on $30 \%$ of the sites. 
$\chi$ has the same cusp, as in Fig.10, flattening at low temperature, 
probably indicative
of spin glass freezing. We are studying the spin glass order
parameter for this problem, as well as the effect of 
on site disorder. There has been some work on RKKY spin glasses
in three dimension \cite{bray,chandan,matsubara} but the 
results about a spin glass transition (in the absence of
anisotropy) are still 
inconclusive.
There is no work, to the extent we know, on models with site dilution 
as well as disorder.

\section{Conclusions}

We wanted to explore the interplay of structural disorder
and doped magnetic moments in an electron system within a
simple model. 
Though we are far from demonstrating the experimentally
observed features near the i-m-t, or detailed properties
of the observed spin glass, we have highlighted how
strong disorder can generate most of these effects.
For instance, relatively weak `spin disorder' ({\it i.e}
small $J'$), acting {\it on top of} structural 
disorder, can have remarkable effects on transport.
This spin disorder acting on its own, in a high
density electron system, would have led only to
weak scattering, masked by electron-phonon effects etc.
Here the change $\Delta \rho$ is large, and so is
the MR. Similarly, the 
occurence of a {\it disordered} magnetic
state, which is the key to the i-m-t and MR, is
again facilitated by structural disorder. We saw how
this happens, due to frustration, 
for a periodic array of spins.
We have not shown results on the actual temperature
and field dependence of transport, we will
discuss these separately \cite{gdsipap}.
Most importantly, we have not provided a hint of
how Coulomb effects can be incorporated.
Let us provide a qualitative discussion.

A controlled perturbative scheme \cite{altaron}
 for handling e-e effects
exists at weak disorder.
This should allow an understanding of the metallic phase, far
from the transition, {\it in terms of the spectrum and 
wavefunctions computed in our scheme}. Similarly, deep in the insulator,
a knowledge of the localisation length and screening 
allows us to quantify the
effects of e-e interactions \cite{efros}.
There is no detailed  theory which allows us to access
physical properties near
the i-m-t (see the work \cite{mottand}
by
V. Dobrosavljevic and G. Kotliar though), although there are
RG calculations (reviewed \cite{belkirk} by Belitz and Kirkpatrick).
There is also an interesting scheme \cite{sadov}
suggested by 
Kuchinskii {\it et al.} but the `control' in their
procedure is not obvious.
So, although it seems difficult to start completely from
first principles, including e-e and electron-spin
interactions, an understanding
of our simplified model can help in quantifying the e-e
effects in the problem.

\section{Acknowledgment}
One of us (P.M) would like to thank T. V. Ramakrishnan for 
a discussion, and the Isaac Newton Institute, Cambridge,
 for hospitality
during part of this work.


\begin{thebibliography}{99}
\bibitem{condcalc} 
The Kubo formula for d.c conductivity in the non interacting case is
discussed in Mott and Davis \cite{mottdavis}. There are some problems
in an actual numerical implementation of this scheme because 
the eigenstates
are typically `real' and non degenerate in the cases we consider. In the 
infinite volume limit the states will be quasi-degenerate and the
conductivity (at fixed 
$\epsilon_F$) can be computed from this manifold. In our finite system
we compute $\sigma(\epsilon)$ by  considering 
two successive states in the spectrum.
\end{thebibliography}
\end{document}